\documentclass[aps,physrev,twocolumn,superscriptaddress]{revtex4-1}

\usepackage[pdftex]{graphicx}
\usepackage{dcolumn}
\usepackage{bm}
\usepackage{amsmath, amssymb,amsfonts}
\usepackage{mathrsfs}
\usepackage{type1cm}
\usepackage[colorlinks,linkcolor=blue,citecolor=blue]{hyperref}

\begin{document}

\title{Squeezed-field path-integral description of BCS superconductors}

\author{Kazuma Nagao}
\email{knagao@physnet.uni-hamburg.de}
\affiliation{%
Zentrum f\"ur Optische Quantentechnologien and Institut f\"ur Laserphysik, Universit\"at Hamburg, 22761 Hamburg, Germany\\
}%
\affiliation{%
The Hamburg Center for Ultrafast Imaging, Luruper Chaussee 149, 22761 Hamburg, Germany
}%
\author{Dapeng Li}
\affiliation{%
Zentrum f\"ur Optische Quantentechnologien and Institut f\"ur Laserphysik, Universit\"at Hamburg, 22761 Hamburg, Germany\\
}%
\affiliation{%
The Hamburg Center for Ultrafast Imaging, Luruper Chaussee 149, 22761 Hamburg, Germany
}%
\author{Ludwig Mathey}
\affiliation{%
Zentrum f\"ur Optische Quantentechnologien and Institut f\"ur Laserphysik, Universit\"at Hamburg, 22761 Hamburg, Germany\\
}%
\affiliation{%
The Hamburg Center for Ultrafast Imaging, Luruper Chaussee 149, 22761 Hamburg, Germany
}%

\date{\today}

\begin{abstract}

We develop a squeezed-field path-integral representation for BCS superconductors utilizing a generalized completeness relation of squeezed-fermionic coherent states.
We derive a Grassmann path integral of fermionic quasiparticles that explicitly includes the collective degrees of freedom of the order-parameter dynamics governed by the classical Anderson pseudospin model.
Based on this method, we analyze the spectral function of the single-particle excitations, and show that the squeezed-field path integral for the BCS Hamiltonian describes a bosonic sideband branch that corresponds to the Higgs amplitude mode of BCS superconductors, in addition to reproducing the quasiparticle and quasihole excitation branches described by the BCS mean-field approximation.

\end{abstract}

\maketitle

\section{Introduction} \label{sec: intro}

Superconductors are macroscopic quantum states of matter, described by order parameters of spontaneously-broken symmetries~\cite{anderson2018basic,leggett2006quantum,wen2004quantum}.
The phenomenological mean-field model for superconductors has been first developed by Ginzburg and Landau \cite{leggett2006quantum}, giving a universal description of the thermodynamic phases.
In 1957, Bardeen, Cooper, and Schrieffer (BCS) provided a microscopic description of conventional $s$-wave superconductors~\cite{schrieffer2018theory,leggett2006quantum,bardeen1957theory}, establishing key notions, such as Cooper pairing of electrons, the phonon-mediated pairing mechanism of Cooper pairs due to lattice vibrations, and the instability of the Fermi surface due to attractive interactions.

Fermionic quasiparticle excitations and bosonic collective excitations are key features of quantum liquids of superfluid fermions exemplified by superconductors~\cite{schrieffer2018theory,leggett2006quantum}, and by liquid $^{3}{\rm He}$~\cite{leggett2006quantum,volovik2003universe}.
Recent pump-probe experiments have explored light-induced dynamics of conventional BCS and high-$T_{c}$ cuprate superconductors near and far-from equilibrium~\cite{shimano2020higgs,pekker2015amplitude,pashkin2010femtosecond,beck2011energy,measson2014amplitude,matsunaga2014light,cea2016nonlinear,krull2016coupling,cea2018polarization,murotani2019nonlinear,giorgianni2019leggett,niwa2019light,chu2020phase}.
A striking direction of research is the exploration of the Higgs amplitude mode in an NbN superconductor with terahertz light~\cite{shimano2020higgs,matsunaga2014light}.
Collective and single-particle properties of superfluid fermions have also been studied in ultracold gases~\cite{giorgini2008theory,tsuchiya2009single,schneider2010universal,gaebler2010an,watanabe2013low,hannibal2015quench,kemper2015direct,hoinka2017bragg,behrle2018higgs,mukherjee2019spectral,loon2020beyond,patel2020universal}.

For a complete understanding of the experimental observations of fermionic superfluid systems, a microscopic description for the co-existence and interaction of the single-particle and collective degrees of freedom has to be established.
For example, quasiparticle pair-breaking processes of Cooper pairs have a significant influence on the collective excitation branch of superconductors and superfluid Fermi gases~\cite{cea2016nonlinear,kurkjian2019pair}.
Despite considerable theoretical advances, a unifying framework of treating both quasiparticle and collective degrees of freedom has not been established so far.

In this paper, we develop a generalized Grassmann path-integral approach to describe the fermionic quasi-particles and the bosonic collective excitations on equal footing.
Our formulation presented in this work is based on the squeezed-field path-integral description, which has been developed in the context of Bose--Einstein condensates of ultracold gases~\cite{seifie2019squeezed}, and more recently, extended to one-dimensional gapless systems described by Tomonaga--Luttinger liquid theory~\cite{nagao2020fluctuations}.
The squeezed-field path integrals are a field theoretical representation of quantum systems, in which one inserts the completeness relation of the squeezed coherent states, not the coherent states, into the time evolutions of the path integral.
The constructed path integrals have an extended phase space of the squeezing parameters for quasiparticles, which include quantum and thermal fluctuations that are hard to access in the standard coherent-state representation.
See Sec.~\ref{sec: method} for a detailed discussion.

Intuitively, the squeezed-field path-integral method is a method that advances the parameters of the canonical Bogoliubov transformation~\cite{ibach2003solid,fetter2012quantum,anderson1958random} to quantized bosonic field variables coupling to the Bogoliubov quasiparticle fields~\cite{seifie2019squeezed}.
Such dynamical squeezing parameters are referred to as squeezing fields~\cite{seifie2019squeezed,nagao2020fluctuations}. 
The Bogoliubov transformation plays an essential role in the mean-field description of the ground state of BCS superconductors, and the description of the quasiparticle excitation spectrum.
The standard mean-field description is represented by a quadratic mean-field Hamiltonian with a finite gap function, which will be presented in Eq.~(\ref{eq: quadratic_hamiltonian_nondiagonal}), and it is diagonalized by the Bogoliubov transformation for a set of constant squeezing parameters.
However, such constant parameters imply a constant gap function, therefore the fluctuations of the order parameter are not present in the description.
As we will see below, our formulation, developed specifically for BCS superconductor systems, allows to write down a general field-theoretical action that describes the field configurations of the fermionic quasiparticle fields and the bosonic squeezing fields.
In particular, the fluctuations of the squeezing fields express the motion of the order parameter around a configuration corresponding to the BCS ground state.
We demonstrate that the fluctuations of the squeezing fields are described by the classical Anderson pseudospin model~\cite{anderson1958random,tsuji2015theory,matsunaga2014light}, whose Hamiltonian is given by an energy functional of electrons with respect to the BCS variational state.
Note that the known mean-field result of the quasiparticle spectrum can be naturally recovered by an approximation of our generalized action, in which the squeezing fields have a mean-field configuration corresponding to the BCS ground state and do not fluctuate. 
Additionally, we analyze experimentally-relevant quantities, in particular, the single-particle spectral function, based on the method developed here, and demonstrate that the squeezed-field path integral produces a bosonic sideband dispersion relation in this quantity, in addition to the fermionic quasiparticle dispersion relation of the BCS mean-field theory.  
We discuss that the sideband dispersion shows key features of the Higgs modes of BCS superconductors in the weakly-interacting BCS regime.
The appearance of the sideband peak in the spectral function has not been discussed in the previous studies about the collective excitations on superconductors, see e.g. Refs.~\cite{pekker2015amplitude,anderson1958random,bogoljubov1958new}.
Finally, we note that our formalism can be readily applied to more complex orders as well, such as unconventional $p$-wave superconductors and superfluids~\cite{mackenzie2003the,volovik2003universe}, and strongly-correlated Hubbard-type systems~\cite{hofstetter2018quantum}.

This paper is organized as follows:
In Sec.~\ref{sec: model}, we introduce the BCS Hamiltonian for electrons interacting via attractive interactions and explain the assumptions for the analysis of the following sections. 
In Sec.~\ref{sec: mfa}, we present a brief overview of the mean-field properties of the BCS Hamiltonian.
In Sec.~\ref{sec: method}, we derive a generalized Grassmann-field path-integral representation of the BCS Hamiltonian by utilizing the completeness relation of squeezed-fermionic coherent states, and give a physical interpretation of the motion of the squeezing field.
In Sec.~\ref{sec: spectrum}, we analyze the energy spectrum of the linearized approximation of the Anderson pseudospin model.
In Sec.~\ref{sec: applications}, we apply the formalism to the single-particle spectral function. 
Finally, in Sec.~\ref{sec: conclusions}, we conclude.

\section{Model} \label{sec: model}

We consider a system of electrons with an attractive interaction, described by the BCS Hamiltonian~\cite{altland2010condensed,bardeen1957theory}
\begin{align}
{\hat {\cal H}} &= \sum_{{\bm k},\sigma=\uparrow,\downarrow}\xi_{\bm k}{\hat c}^{\dagger}_{{\bm k},\sigma}{\hat c}_{{\bm k},\sigma} - \frac{g}{V}\sum_{{\bm k},{\bm k}',{\bm q}}{\hat c}^{\dagger}_{{\bm k}+{\bm q},\uparrow}{\hat c}^{\dagger}_{-{\bm k},\downarrow}{\hat c}_{-{\bm k}'+{\bm q},\downarrow}{\hat c}_{{\bm k}',\uparrow}, \label{eq: bcs_hamiltonian} 
\end{align}
where $\{{\hat c}_{{\bm k},\sigma},{\hat c}^{\dagger}_{{\bm k}',\sigma'}\}={\hat c}_{{\bm k},\sigma}{\hat c}^{\dagger}_{{\bm k}',\sigma'}+{\hat c}^{\dagger}_{{\bm k}',\sigma'}{\hat c}_{{\bm k},\sigma}=\delta_{{\bm k},{\bm k}'}\delta_{\sigma,\sigma'}$, $\mu$ is the chemical potential, $g>0$ is the attractive interaction strength, and $V$ is the total volume.
The free-particle dispersion $\xi_{\bm k}=\epsilon_{\bm k}-\mu$ is measured relative to the chemical potential.
Throughout this work we assume a spherical band structure, i.e. $\epsilon_{\bm k}=\frac{\hbar^2 {\bm k}^2}{2m}$, where $m$ is the electron mass.
Furthermore, we assume that the interaction range is restricted inside an energy region around the free Fermi energy $E_{\rm F} \equiv \frac{\hbar^2k^2_{\rm F}}{2m}$ with a width $2\hbar\omega_{\rm D}$~\cite{ibach2003solid,altland2010condensed}. 
$\omega_{\rm D}$ is the Debye frequency of solids.
We assume that the system is in homogeneous three-dimensional space.

In real solids, electrons couple to the electromagnetic field, which leads to gapped plasmon modes~\cite{altland2010condensed,homann2020higgs}. 
This coupling is ignored in the BCS Hamiltonian~(\ref{eq: bcs_hamiltonian}).
As a consequence, the BCS Hamiltonian possesses a Nambu--Goldstone (NG) mode~\cite{altland2010condensed,nambu1960quasi,goldstone1961field} as a gapless excitation branch with spontaneously-broken U(1) symmetry.
However, this paper is aimed at demonstrating our theoretical formalism.
Hence we ignore the coupling to the gauge fields in the following discussion, to be included elsewhere.

\section{Mean-field properties of the BCS Hamiltonian} \label{sec: mfa} 

Before proceeding, we review the BCS variational ansatz for the ground state of Eq.~(\ref{eq: bcs_hamiltonian}), and the Bogoliubov approximation to obtain the fermionic quasiparticles excitations of the mean-field ground state.
The mean-field properties of the ground state of Eq.~(\ref{eq: bcs_hamiltonian}) are derived from the BCS variational state~\cite{altland2010condensed,fradkin2013field,bardeen1957theory,anderson1958random}:
\begin{align}
|\Omega_{\rm BCS}\rangle = \prod_{\bm k}\left[ u_{\bm k} + v_{\bm k}{\hat c}^{\dagger}_{{\bm k}\uparrow}{\hat c}^{\dagger}_{-{\bm k}\downarrow}\right]|0\rangle, \label{eq: bcs_state}
\end{align}
where ${\bm k}$ runs over momentum space and $|0\rangle$ is the vacuum of fermions.
The variational parameters $u_{\bm k}$ and $v_{\bm k}$ represent the occupation probability of the states $|0_{\bm k}0_{-{\bm k}}\rangle$ and $|\uparrow_{{\bm k}}\downarrow_{-{\bm k}}\rangle$.
The normalization of the BCS state implies a constraint for a spinor $(u_{\bm k},v_{\bm k})^{\rm T}$, i.e. $|u_{\bm k}|^2 + |v_{\bm k}|^2=1$.

The expectation value of the Hamiltonian, i.e. $E_{\rm BCS} = \langle \Omega_{\rm BCS}|{\hat {\cal H}}|\Omega_{\rm BCS}\rangle$, gives a variational functional with respect to the independent parameter $v_{\bm k}$.
From the variational condition $\delta E_{\rm BCS} / \delta v_{\bm k} = \delta E_{\rm BCS} / \delta v^*_{\bm k} = 0$, the equilibrium values of ${u}_{\bm k}$ and ${v}_{\bm k}$ are determined as \cite{altland2010condensed,ibach2003solid}
\begin{align}
{\overline u}^2_{\bm k} 
&= \frac{1}{2}\left(1+\frac{\xi_{\bm k}-gn_0}{E_{\bm k}}\right), \nonumber \\
{\overline v}^2_{\bm k} 
&= \frac{1}{2}\left(1-\frac{\xi_{\bm k}-gn_0}{E_{\bm k}}\right). \label{eq: equilibrium_bogoliubov}
\end{align}
The function $\Delta = \frac{g}{V}\sum_{\bm k}\langle {\hat c}_{-{\bm k},\downarrow}{\hat c}_{{\bm k},\uparrow} \rangle = \frac{g}{V} \sum_{\bm k}{\overline u}_{\bm k}{\overline v}_{\bm k}$ is the gap function of the superconducting phase, and $E_{\bm k}=\sqrt{\Delta^2+(\xi_{\bm k}-gn_0)^2}$.
In addition, $gn_0 \equiv \frac{g}{V}\sum_{\bm k}{\overline v}^2_{\bm k}$ denotes the Hartree shift to the bare chemical potential $\mu$~\cite{micnas1990superconductivity}.
The value of the gap function is evaluated by solving the self-consistent gap equation~\cite{ibach2003solid}.
If the interaction strength $g$ is weak, i.e., $gN_{\rm F} \ll 1$, the gap function is approximately given by $\Delta \approx 2\hbar\omega_{\rm D}\exp\left(-1/(gN_{\rm F})\right) \equiv \Delta'$ with $\mu' \equiv \mu + gn_0 \approx E_{\rm F}$~\cite{micnas1990superconductivity,ibach2003solid}.
Here $N_{\rm F}$ denotes the density of states per Cooper pair and per unit volume on the Fermi surface.
For a three-dimensional system, $N_{\rm F}$ is given by $N_{\rm F}=\frac{(2m)^{3/2}}{4\pi^2\hbar^3}\sqrt{E_{\rm F}}$~\cite{ibach2003solid}.

The Bogoliubov mean-field approximation performs a subtraction of the pairing operators ${\hat c}_{-{\bm k},\downarrow}{\hat c}_{{\bm k},\uparrow}$ included in the two-body interaction term, as ${\hat c}_{-{\bm k},\downarrow}{\hat c}_{{\bm k},\uparrow} = \langle {\hat c}_{-{\bm k},\downarrow}{\hat c}_{{\bm k},\uparrow} \rangle + ({\hat c}_{-{\bm k},\downarrow}{\hat c}_{{\bm k},\uparrow}-\langle {\hat c}_{-{\bm k},\downarrow}{\hat c}_{{\bm k},\uparrow} \rangle)$~\cite{altland2010condensed}. 
The subtracted operator ${\hat c}_{-{\bm k},\downarrow}{\hat c}_{{\bm k},\uparrow}-\langle {\hat c}_{-{\bm k},\downarrow}{\hat c}_{{\bm k},\uparrow} \rangle$ is assumed to be a small fluctuation around the mean-field ground state. 
Note that the expectation values of the pairing operators are taken for the BCS state of Eq.~(\ref{eq: bcs_state}).
Ignoring higher order fluctuations reduces the many-body Hamiltonian (\ref{eq: bcs_hamiltonian}) to a quadratic form, i.e.
\begin{align}
{\hat {\cal H}} &\approx \frac{V|\Delta|^2}{g} + \sum_{\bm k}\xi_{\bm k}({\hat c}^{\dagger}_{{\bm k},\uparrow}{\hat c}_{{\bm k},\uparrow}+{\hat c}^{\dagger}_{-{\bm k},\downarrow}{\hat c}_{-{\bm k},\downarrow})  \nonumber \\
&\;\;\;\; - \sum_{\bm k}\left( \Delta^*{\hat c}_{-{\bm k},\downarrow}{\hat c}_{{\bm k},\uparrow} + \Delta {\hat c}^{\dagger}_{{\bm k},\uparrow}{\hat c}^{\dagger}_{-{\bm k},\downarrow} \right). \label{eq: quadratic_hamiltonian_nondiagonal}
\end{align}
This reduced Hamiltonian is readily diagonalized by using the Bogoliubov transformation~\cite{ibach2003solid,altland2010condensed,fetter2012quantum,fradkin2013field}, defined by   
\begin{align}
{\hat c}_{{\bm k},\uparrow}
&= u_{\bm k}{\hat \psi}_{{\bm k},\uparrow} + v_{\bm k}{\hat \psi}^{\dagger}_{-{\bm k},\downarrow}, \nonumber \\
{\hat c}_{-{\bm k},\downarrow} 
&= u_{\bm k}{\hat \psi}_{-{\bm k},\downarrow} - v_{\bm k}{\hat \psi}^{\dagger}_{{\bm k},\uparrow}. \label{eq: bogoliubov_transformation}
\end{align}
The new field operators ${\hat \psi}^{\dagger}_{{\bm k},\sigma}$ and ${\hat \psi}_{{\bm k},\sigma}$ are the creation and annihilation operators of the Bogoliubov quasiparticles on the BCS ground state.
If we fix $u_{\bm k}$ and $v_{\bm k}$ so that off-diagonal terms vanish, Eq.~(\ref{eq: quadratic_hamiltonian_nondiagonal}) becomes a diagonal form as ${\hat {\cal H}} \rightarrow \sum_{\bm k}E_{\bm k}({\hat \psi}^{\dagger}_{{\bm k},\uparrow} {\hat \psi}_{{\bm k},\uparrow} + {\hat \psi}^{\dagger}_{-{\bm k},\downarrow}{\hat \psi}_{-{\bm k},\downarrow})$, and the BCS state is then found to be the vacuum state of the fermionic quasiparticles.
Note that the Bogoliubov approximation based on the mean-field subtraction does not take into account the Hartree shift to the chemical potential.
Therefore, the diagonal Hamiltonian corresponds to ${\overline u}_{\bm k}$ and ${\overline v}_{\bm k}$ of Eq.~(\ref{eq: equilibrium_bogoliubov}) with no Hartree shift correction.
The energy $E_{\bm k}$ of the diagonal Hamiltonian, which also appears in Eq.~(\ref{eq: equilibrium_bogoliubov}), represents the dispersion relation of the single quasiparticle and quasihole excitations of the BCS state.

In the above mean-field description of the system, the parameters of the Bogoliubov transformation $u_{\bm k}$ and $v_{\bm k}$ have constant values.
Therefore, the gap function, which is associated with the condensate component of the ground state, is a constant. 
The approximate energy spectrum of the many-body system describes only one set of the fermionic quasiparticle branches.
However, in the squeezed-field path integral formulation, as already mentioned in Sec.~\ref{sec: intro}, the parameters become the squeezing fields. 
The squeezing fields are quantized in the sense of the path integral, and their fluctuations relative to a mean-field configuration correspond to the fluctuations of the superconducting order in the system of Eq.~(\ref{eq: bcs_hamiltonian}), see also Sec.~\ref{sec: method}.

\section{Squeezed field path integral for the BCS Hamiltonian} \label{sec: method}

The key ingredient to build a squeezed-field path integral for fermions is the two-mode squeezing operator to create the BCS state:
\begin{align}
{\hat S}({\bm \Omega}) = \prod_{\bm k}\exp\left(\eta_{\bm k}{\hat c}^{\dagger}_{{\bm k},\uparrow}{\hat c}^{\dagger}_{-{\bm k},\downarrow}-\eta^*_{\bm k}{\hat c}_{-{\bm k},\downarrow}{\hat c}_{{\bm k},\uparrow}\right),
\end{align}
where $\eta_{\bm k}=\frac{\theta_{\bm k}}{2}e^{i\varphi_{\bm k}}$ and ${\bm \Omega}=(\cdots,\theta_{\bm k},\varphi_{\bm k},\cdots)$ is a vector of the parameters of the squeezing transformation.
This definition implies a parametrization of $u_{\bm k}$ and $v_{\bm k}$ in Eq.~(\ref{eq: bogoliubov_transformation}) as $u_{\bm k}=\cos(\theta_{\bm k}/2)$ and $v_{\bm k} = \sin(\theta_{\bm k}/2)e^{i\varphi_{\bm k}}$.
We choose a specific gauge of $\theta_{\bm k}$ and $\varphi_{\bm k}$ as $ 0 \leq \theta_{\bm k} \leq \pi$ and $ 0 \leq \varphi_{\bm k} < 2\pi$.
Due to the unitarity of the transformation, the identity ${\hat S}^{\dagger}({\bm \Omega}){\hat S}({\bm \Omega})={\hat S}({\bm \Omega}){\hat S}^{\dagger}({\bm \Omega})=1$ is satisfied for arbitrary $\eta_{\bm k}$.
In terms of this unitary operator, the BCS state is written as a squeezed vacuum state of electrons, i.e. $|\Omega_{\rm BCS}\rangle = {\hat S}({\bm \Omega})|0\rangle$.
See also, e.g., Refs.~\cite{anderson1958random,read2000paired,fradkin2013field}.

In the standard path-integral method for fermions, the fermionic coherent states of Grassmann numbers are used to span the phase space for classical trajectories \cite{altland2010condensed,fradkin2013field}.
These are defined as a ket vector $|{\bm c}\rangle = e^{{\hat {\bm c}}^{\dagger}\cdot {\bm c} - {\overline {\bm c}}\cdot{\hat {\bm c}}}|0\rangle$, and its conjugated bra is $\langle {\overline {\bm c}} | = \langle 0 |e^{-{\hat {\bm c}}^{\dagger}\cdot {\bm c} + {\overline {\bm c}}\cdot{\hat {\bm c}}}$.
The vectors, ${\bm c}=(\cdots,c_{{\bm k},\sigma},\cdots)$ and ${\bm {\overline c}}=(\cdots,{\overline c}_{{\bm k},\sigma},\cdots)$, are the Grassmann-number fields, and $\langle {\overline {\bm c}}|{\bm c}\rangle = 1$.
Similar to the bosonic coherent state, the fermionic coherent state satisfies a completeness relation \cite{altland2010condensed,fradkin2013field}, given by
\begin{align}
\int d{\overline {\bm c}}d{\bm c} \; |{\bm c}\rangle \langle {\overline {\bm c}} | = 1. \label{eq: complete_fermi}
\end{align}
The integral measure is $d{\overline {\bm c}}d{\bm c} = \prod_{\bm k}d{\overline c}_{\bm k}dc_{\bm k}$.
This completeness relation is used to derive a classical action of the Grassmann fields for the quantum-mechanical Hamiltonian (\ref{eq: bcs_hamiltonian}).
The Grassmann-field representation has been widely utilized to formulate perturbative and non-perturbative frameworks for many-body problems of interacting fermions~\cite{altland2010condensed,auerbach2012interacting,fradkin2013field,shankar1994renormalization}.
For example, applications of the Grassmann path integrals to renormalization group analyses of interacting fermions have been comprehensively reviewed in~\cite{shankar1994renormalization}.

To obtain a squeezed-field path integral for ${\hat {\cal H}}$, we first squeeze the completeness relation (\ref{eq: complete_fermi}) with ${\hat S}({\bm \Omega})$, and then integrate over it with respect to the variational parameters $\eta_{\bm k}$.
The ${\bm \Omega}$-integration is normalized by using the Haar invariant measure of the SU(2) group, i.e. $d{\bm \Omega} = \prod_{\bm k}\frac{d\theta_{\bm k} d\varphi_{\bm k}}{4\pi} \sin \theta_{\bm k}$~\cite{auerbach2012interacting,fradkin2013field,note1}.
As a result, we arrive at the relation 
\begin{align}
\int d{\bm \Omega} d{\overline {\bm \psi}}d{\bm \psi} \; {\hat S}({\bm \Omega})|{\bm \psi}\rangle \langle {\overline {\bm \psi}} |{\hat S}^{\dagger}({\bm \Omega}) = 1. \label{eq: general_completeness}
\end{align}
Here we use ${\bm \psi}=(\cdots,\psi_{{\bm k},\sigma},\cdots)$ in order to emphasize that the Grassmann fields in Eq.~(\ref{eq: general_completeness}) represent the Bogoliubov quasiparticle fields.
This equation (\ref{eq: general_completeness}) provides an extended completeness relation for the squeezed-fermionic coherent state $|{\bm \Omega},{\bm \psi}\rangle \equiv {\hat S}({\bm \Omega})|{\bm \psi}\rangle$.
In the path integrals built with the squeezed-fermionic coherent states, the classical trajectories move through the extended phase space $({\bm \psi},{\overline {\bm \psi}},{\bm \Omega})$, which includes the phase space $({\bm \psi},{\overline {\bm \psi}})$ of Eq.~(\ref{eq: complete_fermi}) as its subspace.
The additional motion along the direction of ${\bm \Omega}$ includes additional fluctuations around the BCS state, which are not easily accessible in the standard phase-space choice of the Grassmann path integrals, i.e., Eq.~(\ref{eq: complete_fermi}).

Consider the thermodynamic partition function for the BCS Hamiltonian $Z(\beta,\mu)={\rm Tr} e^{-\beta{\hat {\cal H}}}$, where $\beta = (k_{\rm B}T)^{-1}$ is the inverse temperature.
We insert the extended completeness relation of the squeezed-fermionic coherent state and take the continuum limit to obtain 
\begin{align}
Z = \int_{0\leq |v_{\bm k}|^2 \leq 1} {\cal D}({\overline {\bm \psi}},{\bm \psi},{\bm v}^*,{\bm v})e^{-\frac{1}{\hbar}{\cal S}({\overline {\bm \psi}},{\bm \psi},{\bm v}^*,{\bm v})}. \label{eq: definition_partition}
\end{align}
We impose the constraint condition, $0\leq |v_{\bm k}|^2 \leq 1$ for each ${\bm k}$ on the path integral to be consistent with the integration area of the Haar measure.
The Euclidean action ${\cal S}=\int^{\hbar \beta}_{0} d\tau {\cal L}({\overline {\bm \psi}},{\bm \psi},{\bm v}^*,{\bm v})$ is the time integral of the squeezed-field Lagrangian 
\begin{align}
{\cal L}({\overline {\bm \psi}},{\bm \psi},{\bm v}^*,{\bm v}) &= \sum_{{\bm k},\sigma}{\overline \psi}_{{\bm k},\sigma}\hbar\partial_{\tau}\psi_{{\bm k},\sigma} + \sum_{\bm k} v^*_{\bm k}\hbar\partial_{\tau}v_{\bm k} \nonumber \\
&\;\;\;\;\;\;\;\;\;\;\;\; + H({\overline {\bm \psi}},{\bm \psi},{\bm v}^*,{\bm v}) + {\cal L}_{\rm NLD}.
\end{align}
The classical Hamiltonian $H({\overline {\bm \psi}},{\bm \psi},{\bm v}^*,{\bm v}) = \langle {\bm {\overline \psi}}, {\bm \Omega}| {\hat {\cal H}} | {\bm \Omega}, {\bm \psi} \rangle$ is the expectation value of the BCS Hamiltonian ${\hat {\cal H}}$ with respect to the squeezed fermionic coherent state $|{\bm \Omega}, {\bm \psi} \rangle$.
The dynamical term, ${\cal L}_{\rm dyn} \equiv \sum_{{\bm k},\sigma} {\overline \psi}_{{\bm k},\sigma}\hbar\partial_{\tau} {\psi}_{{\bm k},\sigma} + \sum_{\bm k} v^*_{\bm k}\hbar\partial_{\tau}v_{\bm k}  + {\cal L}_{\rm NLD}$, is the Berry-phase term of the imaginary-time path integral, which stems from the continuum limit of the direct product of the adjacent overlaps $\prod_{j=1}^{\infty} \langle {\overline {\bm \psi}}_{j},{\bm \Omega}_{j} | {\bm \Omega}_{j-1},{\bm \psi}_{j-1}\rangle $~\cite{auerbach2012interacting}.
See also Appendix~\ref{app: evaluation_overlap} for the derivation.
The first term ${\cal L}_{\rm dyn,1} \equiv \sum_{{\bm k},\sigma} {\overline \psi}_{{\bm k},\sigma} \hbar\partial_{\tau} {\psi}_{{\bm k},\sigma}$ is the dynamical term of the standard Grassmann path integrals~\cite{altland2010condensed}.
The second dynamical term, ${\cal L}_{\rm dyn,2} \equiv \sum_{\bm k} v^*_{\bm k}\hbar\partial_{\tau}v_{\bm k}$, describes the dynamics of the squeezing fields $v_{\bm k}$ and $v^*_{\bm k}$.
Furthermore, as an additional feature of squeezed-field path integrals, ${\cal L}_{\rm dyn}$ includes a nonlinear dynamical term~\cite{seifie2019squeezed,nagao2020fluctuations}, given by
 \begin{align}
{\cal L}_{\rm NLD} 
&=  \sum_{\bm k}\frac{\hbar{\dot \theta}_{{\bm k}}}{2}\left( e^{i\varphi_{{\bm k}}}{\overline \psi}_{{\bm k},\uparrow}{\overline \psi}_{-{\bm k},\downarrow} - e^{-i\varphi_{{\bm k}}}\psi_{-{\bm k},\downarrow}\psi_{{\bm k},\uparrow} \right) \nonumber \\
&-\sum_{\bm k} i \hbar {\dot \varphi}_{{\bm k}} |v_{{\bm k}}|^2 \left( {\overline \psi}_{{\bm k},\uparrow}\psi_{{\bm k},\uparrow} + {\overline \psi}_{-{\bm k},\downarrow}\psi_{-{\bm k},\downarrow} \right) \label{eq: lagrangian_nld} \\
&+\sum_{\bm k} i \hbar {\dot \varphi}_{{\bm k}}\left(u_{{\bm k}}v_{{\bm k}} {\overline \psi}_{{\bm k},\uparrow}{\overline \psi}_{-{\bm k},\downarrow} + u_{{\bm k}}v^*_{{\bm k}} \psi_{-{\bm k},\downarrow}\psi_{{\bm k},\uparrow} \right). \nonumber
\end{align}
A similar set of dynamical terms also appears in squeezed-field path integrals for bosonic systems, see Refs.~\cite{seifie2019squeezed,nagao2020fluctuations}.

The classical Hamiltonian $H$ can be expressed as $H[{\overline {\bm \psi}},{\bm \psi},{\bm v}^*,{\bm v}] = E_{\rm BCS}[{\bm v}^*,{\bm v}] + \langle {\overline {\bm \psi}} | : {\hat S}^{\dagger}({\bm \Omega}){\hat {\cal H}}{\hat S}({\bm \Omega}) : | {\bm \psi} \rangle$.
The double colons represent the normal ordering operation for the Bogoliubov-transformed fermions.
The expression of the energy functional $E_{\rm BCS}$ is \cite{anderson1958random}
\begin{align}
E_{\rm BCS}[{\bm v}^*,{\bm v}] &= \sum_{\bm k}2\xi_{\bm k}|v_{\bm k}|^2 - \frac{g}{V}\sum_{{\bm k}_1,{\bm k}_2}u_{{\bm k}_1}v_{{\bm k}_1}u_{{\bm k}_2}v^*_{{\bm k}_2} \nonumber \\
&\;\;\;\;\;\;\;\;\;\;\;\;\;\; - \frac{g}{V}\sum_{{\bm k}_1,{\bm k}_2}|v_{{\bm k}_1}|^2|v_{{\bm k}_2}|^2. \label{eq: pseudospin_model}
\end{align}
To advance our analysis of this classical model, we introduce Anderson's pseudospins for the BCS state~\cite{anderson1958random,tsuji2015theory,matsunaga2014light}
\begin{align}
s^{x}_{\bm k} &= \frac{1}{2}\sin\theta_{\bm k}\cos \varphi_{\bm k} = \frac{1}{2}u_{\bm k}(v_{\bm k}+v^*_{\bm k}), \nonumber \\
s^{y}_{\bm k} &= \frac{1}{2}\sin\theta_{\bm k}\sin \varphi_{\bm k} = \frac{1}{2i}u_{\bm k}(v_{\bm k}-v^*_{\bm k}),  \\
s^{z}_{\bm k} &= \frac{1}{2}\cos\theta_{\bm k} = \frac{1}{2}(u^2_{\bm k}-|v_{\bm k}|^2). \nonumber
\end{align}
In this spin representation, the energy functional (\ref{eq: pseudospin_model}) becomes a long-range interacting classical-spin model defined on the ${\bm k}$-lattice~\cite{anderson1958random}.
Indeed, the first term of (\ref{eq: pseudospin_model}) translates to a linear term of $s^{z}_{\bm k}$ coupled to an external inhomogeneous {\it magnetic field}, i.e., $\xi_{\bm k}$.
The second term describes the long-range $XY$ couplings between $s^{x,y}_{\bm k}$ and $s^{x,y}_{{\bm k}'}$ with the strength $g/V$, and the last term denotes long-range Ising interactions between $s^{z}_{\bm k}$ and $s^{z}_{{\bm k}'}$.
The interaction range of the pseudospins is over the restricted momentum region around the Fermi surface.
In the squeezed-field path integral for BCS superconductors, the squeezing parameters $v_{\bm k}$ are dynamical fields of the path integral, expressing quantum and thermal fluctuations of the order-parameter field.
In terms of the pseudospins, the dynamics can be visualized as precessing vectors $(s^{x}_{\bm k},s^{y}_{\bm k},s^{z}_{\bm k})$ for each momentum ${\bm k}$~\cite{tsuji2015theory,matsunaga2014light}.
We note that the Anderson pseudospin model has been used to analyze the pump-probe response of the Higgs mode of $s$-wave superconductors~\cite{matsunaga2014light,shimano2020higgs}.

To simplify the squeezed-field Lagrangian, we consider small fluctuations around the mean-field ordered state.
We split the field variable $v_{\bm k}$ into a mean value ${\overline v}_{\bm k}$ and its fluctuations $b_{\bm k}$, i.e. $v_{\bm k} = {\overline v}_{\bm k}+b_{\bm k}$.
The fluctuations $b_{\bm k}$ are assumed to be small, relative to ${\overline u}_{\bm k}$.
We expand the field $u_{\bm k}=\sqrt{1-v^*_{\bm k}v_{\bm k}} = {\overline u}_{\bm k} - \frac{1}{2}\frac{{\overline v}_{\bm k}}{{\overline u}_{\bm k}}(b_{\bm k}+b_{\bm k}^*) + \cdots$ to obtain ${\cal L} \approx \text{const.} + {\cal L}^{(2)}_{\rm F} + {\cal L}^{(2)}_{\rm B} + {\cal L}_{\rm int}({\overline {\bm \psi}},{\bm \psi},{\bm b}^*,{\bm b},{\dot {\bm \theta}},{\dot {\bm \varphi}})$, in which 
\begin{align}
{\cal L}^{(2)}_{\rm F} &= \sum_{{\bm k},\sigma} {\overline \psi}_{{\bm k},\sigma}(\hbar\partial_{\tau}+\hbar\omega_{\bm k}) {\psi}_{{\bm k},\sigma}, \\
{\cal L}^{(2)}_{\rm B} &= \sum_{\bm k}b^*_{\bm k}\hbar\partial_{\tau}b_{\bm k} + \sum_{{\bm k}_1,{\bm k}_2}2A_{{\bm k}_1,{\bm k}_2}b^*_{{\bm k}_1}b_{{\bm k}_2} \nonumber \\
&+\sum_{{\bm k}_1,{\bm k}_2}B_{{\bm k}_1,{\bm k}_2}\left(b_{{\bm k}_1}b_{{\bm k}_2}+b^*_{{\bm k}_1}b^*_{{\bm k}_2}\right). \label{eq: lagrangian_quad_boson}
\end{align}
The matrices $A_{{\bm k}_1,{\bm k}_2}$ and $B_{{\bm k}_1,{\bm k}_2}$ are given in Appendix~\ref{app: bogoliubov}.
The fermionic part of the quadratic Lagrangian, ${\cal L}^{(2)}_{\rm F}$, describes the dispersion relation of the fermionic Bogoliubov modes~$\hbar\omega_{\bm k} = E_{\bm k}$.
Note that ${\cal L}^{(2)}_{\rm F}$ corresponds to the diagonalized form of Eq.~(\ref{eq: quadratic_hamiltonian_nondiagonal}), but now it includes the Hartree correction to the chemical potential. 
The bosonic part, ${\cal L}^{(2)}_{\rm B}$, describes collective modes of the order parameter expressed as a precessing motion of the pseudospins.
Note that the linear terms of the bosons $b_{\bm k}$ vanish due to the variational condition to determine ${\overline v}_{\bm k}$.
Since the Hamiltonian of ${\cal L}^{(2)}_{\rm B}$ is not diagonal with respect to $b_{\bm k}$ and $b_{\bm k}^*$, the bosons with different momenta have a finite correlation with each other.
In Sec.~\ref{sec: spectrum}, we discuss the energy eigenvalues of this classical Hamiltonian in detail.
Moreover, the nonlinear interaction term ${\cal L}_{\rm int}({\overline {\bm \psi}},{\bm \psi},{\bm b}^*,{\bm b},{\dot {\bm \theta}},{\dot {\bm \varphi}})$ describes the couplings between the fermionic and bosonic degrees of freedom as well as the self-interaction of each component.

Before proceeding, we make several remarks on the generalized path integral and the fluctuation expansion.
First, the pseudospin representation is reminiscent of the Holstein--Primakoff (HP) representation of the SU(2) spin operators, see e.g.~\cite{auerbach2012interacting}.
It is defined via
\begin{align}
{\hat S}^{x} = \frac{{\hat a}^{\dagger}{\hat b}+{\hat b}^{\dagger}{\hat a}}{2},\; {\hat S}^{y} = \frac{{\hat a}^{\dagger}{\hat b}-{\hat b}^{\dagger}{\hat a}}{2i},\; {\hat S}^{z} = \frac{{\hat a}^{\dagger}{\hat a}-{\hat b}^{\dagger}{\hat b}}{2},
\end{align}
in which ${\hat a}={\hat a}^{\dagger}=\sqrt{2S-{\hat b}^{\dagger}{\hat b}}$ is assumed, and $S$ is the strength of the SU(2) spins.
The operators ${\hat b}$ and ${\hat b}^{\dagger}$ satisfy $[{\hat b},{\hat b}^{\dagger}]=1$ and $[{\hat b},{\hat b}]=[{\hat b}^{\dagger},{\hat b}^{\dagger}]=0$.
We note that there is a formal correspondence as ${\hat a}\leftrightarrow u_{\bm k}$ and ${\hat b} \leftrightarrow v_{\bm k}$.
In this analogy, the fluctuation expansion that we have presented above can be seen as the HP spin-wave expansion around a mean-field ground state~\cite{auerbach2012interacting,nagao2018response}.
We note that, unlike the usual HP expansion for a ferromagnetic ground state~\cite{auerbach2012interacting}, ${\hat b}$ has the mean value ${\overline v}_{\bm k}$ for the BCS state.
In the squeezed-field path-integral formulation, the field variables $b_{\bm k}$ are quantized, in analogy to the quantized spin waves of a magnetic state.

Having this observation in mind, the nonlinear term ${\cal L}_{\rm int}({\overline {\bm \psi}},{\bm \psi},{\bm b}^*,{\bm b},{\dot {\bm \theta}},{\dot {\bm \varphi}})$ can be regarded as a sum of infinitely-many nonlinear processes involving three, four, and infinite spin waves in ${\bm k}$-space.
Furthermore, in addition to the contributions to ${\cal L}_{\rm int}$ from the Hamiltonian, the nonlinear contributions also derive from ${\cal L}_{\rm NLD}$ of the Berry phase, see Eq.~(\ref{eq: lagrangian_nld}).
We note that the time derivative terms are essential in determining the equations of motion in the Lagrange formalism~\cite{nair2005quantum,kramer2008a}.
However, in the Hamilton formalism, the same equations are derived by introducing Poisson brackets for the canonical phase-space variables, which are independently given of the Hamiltonian~\cite{nair2005quantum,kramer2008a}.
Therefore, the presence of the nonlinear time-derivative terms implies that the Poisson brackets for bosons and fermions are geometrically deformed by such perturbations~\cite{kuratsuji1988deformation}.
We will discuss this consideration more concretely elsewhere. 
We note that a perturbative analysis of the nonlinear interactions of the squeezing fields has been discussed in Ref.~\cite{nagao2020fluctuations}.

In spin-wave theory, corrections due to nonlinear processes are small, if the fluctuations around the reference state of the expansion are sufficiently small.
Otherwise, the truncation of the expansion is no longer valid, see also Ref.~\cite{auerbach2012interacting}.
In Sec.~\ref{sec: applications}, we discuss a parameter regime, in which the corrections due to the fluctuations of the squeezing fields are small compared to the BCS mean-field results.
In the following sections, we focus on properties of the quadratic Lagrangian ${\cal L}^{(2)}_{\rm F} + {\cal L}^{(2)}_{\rm B}$, and visualize the consequences of this quadratic Lagrangian to observable quantities relevant to real systems.

Next, we point out a similarity of the generalized Grassmann path integral to specific features of gauge field theories~\cite{wen2004quantum}.
The path-integral expressions of quantum systems are equivalent representations of the operator-based formalisms.
For the generalized path integral, if we go back to the operator language, we observe an additional state space for the constrained bosons as well as for the fermionic quasiparticles.
Therefore, an additional bosonic Hilbert space has emerged from the original description of the BCS Hamiltonian.
We note that the emergence of the intrinsic dynamics of the squeezing field, i.e., the quantization of $v_{\bm k}$, can be attributed to the mixing property of ${\hat S}({\bm \Omega})$ for particles and holes, which allows to yield multiple operator products, which are not normal ordered in the time evolution of the path integral, see also Appendix~\ref{app: evaluation_overlap}.
Gauge-field descriptions for strongly-correlated systems, such as the $t$-$J$ model and the Hubbard model, have been reviewed in Refs.~\cite{wen2004quantum,lee2006doping,auerbach2012interacting}.
Since the ratio $v_{\bm k}/u_{\bm k}$ or the product $u_{\bm k}v_{\bm k}$ quantify the amount of quantum entanglement between $({\bm k},\uparrow)$ and $(-{\bm k},\downarrow)$ of the BCS state \cite{dunning2005ground,puspus2014entanglement}, in this sense, the bosonic field $v_{\bm k}$ has similarities to a gauge field, which {\it classically} mediates quantum correlations between the fermionic single-particle states.

We note that the construction of a path integral for fermions not based on the standard coherent state is also discussed for the generalized coherent state method~\cite{perelomov1986generalized}, which is based on representation theory of Lie groups.
However, the generalized coherent state method applied to a fermionic system typically needs to project the full Hilbert space into a subspace that arises as an irreducible unitary representation of a Lie group.
For instance, if the system is projected onto a bosonic sector of the Hilbert space $\mathscr{H}_{\rm SU(2)} = \bigotimes_{\bm k}\{|0_{\bm k}0_{-{\bm k}}\rangle,|\uparrow_{\bm k}\downarrow_{-{\bm k}}\rangle\}$, the pairing operators ${\hat l}^{-}_{\bm k}={\hat c}^{\dagger}_{{\bm k},\uparrow}{\hat c}^{\dagger}_{-{\bm k},\downarrow}$ and ${\hat l}^{+}_{\bm k}={\hat c}_{-{\bm k},\downarrow}{\hat c}_{{\bm k},\uparrow}$ and the magnetization ${\hat l}^{z}_{\bm k}=\frac{1}{2}({\hat c}_{-{\bm k},\downarrow}{\hat c}^{\dagger}_{-{\bm k},\downarrow}-{\hat c}^{\dagger}_{{\bm k},\uparrow}{\hat c}_{{\bm k},\uparrow})$ effectively behave as generators of the SU(2) Lie group. 
Then, one can define a generalized coherent state associated with the SU(2) group~\cite{perelomov1986generalized}, which turns out to be the BCS state of Eq.~(\ref{eq: bcs_state}).
In the bosonic sector, the BCS state provides a completeness relation, therefore it generates a bosonic path integral for the fermionic system.
However, the projection is allowed only when the Hamiltonian consists of the SU(2) generators.
Note that the path integral with the SU(2) generalized coherent state is formally obtained by setting $\psi_{{\bm k},\sigma}={\overline \psi}_{{\bm k},\sigma}=0$ in Eq.~(\ref{eq: definition_partition}).
In addition, the squeezed coherent state is not the generalized coherent state defined by a linear combination of Lie generators for a Lie group.
Hence, the squeezed-field Grassmann path integral that we propose in this paper represents a new class of generalized path integrals, which are not simply derived from the group-theoretical method.

\section{Energy spectrum of the linearized pseudospin model} \label{sec: spectrum}

We analyze the energy spectrum of the linearized pseudospin Hamiltonian.
As a preparational step, we introduce an energy-shell representation of the momentum sum, i.e. $\sum_{\bm k} \rightarrow V N_{\rm F}\sum_{i}\Delta \xi$.
We assume that each shell has a spherical shape of area $V N_{\rm F}$ and width $\Delta \xi$.
We note that this replacement is valid only for the $s$-wave systems, which have a spherical Fermi surface, and a spherical pairing symmetry of the Cooper pairs~\cite{leggett2006quantum}.
For this representation, the linearized Hamiltonian of ${\cal L}^{(2)}_{\rm B}$ is formally expressed as
\begin{align}
H^{(2)}_{\rm B} = {\overline {\bm b}} \cdot M \cdot {\bm b}.
\end{align}
The matrix $M$ is the Bogoliubov matrix of $2N$ dimensions, where $N$ is the number of the energy shells.
The notation ${\bm b}=(b_{1},\cdots,b_{N},b^*_{1},\cdots,b^*_{N})^{\rm T}$ is a composite vector of complex-valued fields, and ${\overline{\bm b}}=(b^*_{1},\cdots,b^*_{N},b_{1},\cdots,b_{N})$ is its Hermitian conjugation.

To obtain the energy spectrum of $H^{(2)}_{\rm B}$, we perform the bosonic Bogoliubov transformation~\cite{huber2007dynamical,nagao2018response} for the classical variables, i.e. ${\bm b} = W \cdot {\bm \beta}$.
For later use, especially in Sec.~\ref{sec: applications}, we parametrize the $2N\times 2N$ matrix $W$ as
\begin{align}
W =
\begin{pmatrix}
{\cal U} & {\cal V} \\
{\cal V}^* & {\cal U}^*
\end{pmatrix}
.
\end{align}
The bosonic Bogoliubov transformation has to preserve the Poisson bracket of the bosons, so that the condition 
\begin{align}
W \Sigma_{2N} W^{\rm H} = \Sigma_{2N}, \label{eq: normalization_condition}
\end{align}
is required~\cite{huber2007dynamical,nagao2018response}.
$W^{\rm H}$ represents the Hermitian conjugation of $W$.
The matrix $\Sigma_{2N}=\text{diag}({\mathbf 1}_{N},-{\mathbf 1}_{N})$ is a metric tensor of a $2N$-dimensional Minkowski space.
Hence, the matrix $W$ should be normalized as it is pseudo-unitary.
See references~\cite{huber2007dynamical,nagao2018response} for details and further applications.

\begin{figure}
\begin{center}
\includegraphics[width=85mm]{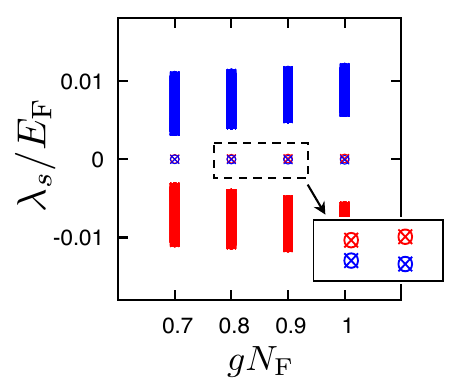}
\vspace{-4mm}
\caption{
Numerical results for the eigenenergies of the squeezing field, at quadratic order, described by ${\cal L}_{\rm B}^{(2)}$, see Eq.~(\ref{eq: lagrangian_quad_boson}). 
This result is obtained via Bogoliubov diagonalization of the matrix $\Sigma_{2N}\cdot M$.
The matrix size is $2N=3200$.
The Debye frequency is $\hbar\omega_{\rm D}=0.01E_{\rm F}$.
The eigenvalues that have positive (negative) norms of the eigenfunctions are expressed with blue (red) points.
We note that we have utilized the DGEEV algorithm of LAPACK for the numerical diagonalization.
(inset): The pairs of the points, which are close to zero, give negative values of $\hbar\omega_{1}^{\rm sq}$.
}
\label{fig: eigenvalue}
\end{center}
\end{figure}

The eigenvalue problem that we solve is $(\Sigma_{2N}M)\cdot W = W \cdot \Lambda$.
The eigenvalue matrix is $\Lambda = \text{diag}(\lambda_1,\cdots,\lambda_{2N})$.
We use the inverse matrix $W^{-1}=\Sigma_{2N}W^{\rm H}\Sigma_{2N}$ to derive
\begin{align}
H^{(2)}_{\rm B} = {\overline {\bm \beta}} \cdot (\Sigma_{2N} \Lambda) \cdot {\bm \beta} = \sum_{s=1}^{N}\hbar\omega_{s}\beta_{s}^*\beta_{s}.
\end{align}
Therefore, the energy eigenvalues of $H^{(2)}_{\rm B}$ are given by $\hbar\omega_{s}^{\rm sq} = \lambda_{s}-\lambda_{s+N}$ for $s=1,\cdots,N$.
We assume that $\lambda_{s} \leq \lambda_{s+1}$ and $ \lambda_{s+N} \geq \lambda_{s+1+N}$ for $s=1,\cdots,N-1$.

We numerically evaluate the eigenvalues $\lambda_{s}$ for $N$ energy shells, ranging from $E_{\rm F}-\hbar\omega_{\rm D}$ to $E_{\rm F}+\hbar\omega_{\rm D}$.
Figure~\ref{fig: eigenvalue} displays the numerical results for $\lambda_{s}$ for $N=1600$, i.e., $2N=3200$, and for several values of the dimensionless interaction strength $gN_{\rm F}$.
For the numerical calculation, we set $\hbar\omega_{\rm D}=0.01 E_{\rm F}$, which corresponds to the typical order of the Debye frequency of BCS superconductors~\cite{ibach2003solid}.
Moreover, the ground-state values of the gap function $\Delta$ are determined by numerically solving the gap equation $\frac{g}{2V}\sum_{\bm k}\frac{1}{E_{\bm k}} = 1$.
This equation is solved for different values of the energy scale $g n_0$, with $n_0 = N_0/V$.
$N_0$ is the particle number inside the interaction regime of the momentum space.
Table~\ref{tab: gaps} presents the calculated values of $\Delta$ for $0.7 \leq gN_{\rm F} \leq 1.0$, and also the approximate values estimated by $\Delta' =  2\hbar\omega_{\rm D}e^{-1/(gN_{\rm F})}$ for reference.
In this regime, we find that the gap function is smaller than the Debye frequency $\hbar\omega_{\rm D}=0.01 E_{\rm F}$, and the chemical potential $\mu'$ is sufficiently close to $1$.
In this work, we only focus on this weakly-interacting regime, i.e., the BCS regime~\cite{chen2005bcs,parish2015bcs}, in three dimensions. 

\begin{figure}
\begin{center}
\includegraphics[width=85mm]{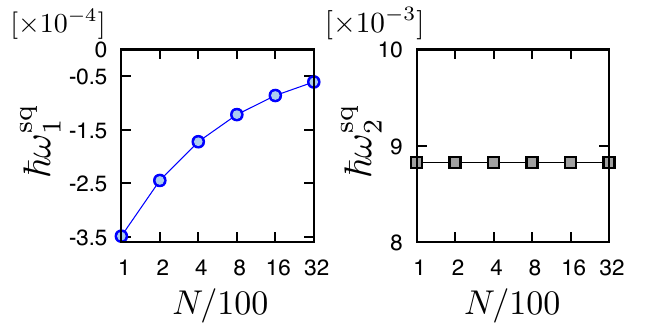}
\vspace{-5mm}
\caption{
(Left): Matrix-size dependence of the lowest eigenvalue $\hbar\omega_{1}^{\rm sq}/E_{\rm F}<0$.
The horizontal axis represents $N$, ranging from $N=100$ to $N=3200$.
The $x$-axis is displayed as log scale with basis $2$.
We put $gN_{\rm F}=0.8$.
(Right): The size dependence of the second eigenvalue $\hbar\omega_{2}^{\rm sq}/E_{\rm F}>0$ for the same $gN_{\rm F}$, implying its convergence.
}
\label{fig: convergence}
\end{center}
\end{figure}

As seen in Fig.~\ref{fig: eigenvalue}, the values of $\hbar\omega^{\rm sq}_{s}$ from $s=2$ to $s=N$ have positive values for all $gN_{\rm F}$.
The energy eigenvalues are continuously distributed over an energy range, for each $gN_{\rm F}$.
As discussed later, this finite-size range of the eigenvalues implies the band width of the Higgs amplitude mode of the BCS superconductors~\cite{pekker2015amplitude,shimano2020higgs}.
In Sec.~\ref{sec: applications}, we will demonstrate that these allowed states form a band dispersion of the Higgs mode in physically-relevant dynamical quantities with a mass gap at the Fermi surface.
We also find that the range of the distribution decreases as $gN_{\rm F}$ increases, implying the reduction of the band width of the Higgs mode with increasing interaction.
We note that the minimum and maximum values of $\hbar\omega_{s>1}^{\rm sq}$, i.e., $\hbar\omega_{2}^{\rm sq}$ and $\hbar\omega_{N}^{\rm sq}$, show good convergence for $N \geq 100$.
The dependence of the second mode $\hbar\omega_{2}^{\rm sq}$ is shown in Fig.~\ref{fig: convergence}.
Moreover, as seen in Fig.~\ref{fig: eigenvalue}, the energy difference between $\hbar\omega_{2}^{\rm sq}$ and $\hbar\omega_{1}^{\rm sq}$ increases with $gN_{\rm F}$, see also Fig.~\ref{fig: convergence}.

\begin{table}[b]
\caption{
Numerically computed values of the gap function.
$\Delta$ denotes the numerical solution of the gap equation, and $\Delta' = 2\hbar\omega_{\rm D}e^{-1/(gN_{\rm F})}$.
The Debye frequency is $\hbar\omega_{\rm D} = 0.01E_{\rm F}$.
}
\begin{ruledtabular}
\begin{tabular}{cccccc}
 &$gN_{\rm F}$ & $\Delta'$ ($E_{\rm F}$) & $\Delta$ ($E_{\rm F}$) \\
\hline
& 0.7 & $4.79 \times 10^{-3}$ & $5.09 \times 10^{-3}$ \\
& 0.8 & $5.73 \times 10^{-3}$ & $6.24 \times 10^{-3}$ \\
& 0.9 & $6.58 \times 10^{-3}$ & $7.38 \times 10^{-3}$ \\
& 1.0 & $7.36 \times 10^{-3}$ & $8.51\times 10^{-3} $ \\
\end{tabular}
\end{ruledtabular}
\label{tab: gaps}
\end{table}

Figure~\ref{fig: eigenvalue} indicates that the eigenvalue for $s=1$, $\hbar\omega_{1}^{\rm sq}$, exhibits small, but finite negative values for all $gN_{\rm F}$.
This mode can be regarded as an onset of the NG mode with zero momentum, i.e., a zero-energy excited state. 
See also Sec.~\ref{sec: applications}.
As shown in Fig.~\ref{fig: convergence}, the finite value of the lowest mode approaches zero as the matrix size $N$ increases.
Therefore, the gap opening of the lowest mode seems to be a finite-size effect, which is similar to the Anderson tower of states of quantum spin chains~\cite{lhuillier2005frustrated,anderson1952an}.
Given this numerical insight, we expect that the energy gap of the lowest mode asymptotically approaches zero from the negative side for larger values of $N$, constituting the NG mode with zero momentum in the thermodynamic limit.

The negativity of the lowest eigenvalue means that the fluctuation expansion of $v_{\bm k}$ up to the quadratic order fails to reproduce the physical properties of the NG modes.
For the negative mode, the Bose distribution function, $f_{\rm B}(\omega^{\rm sq}_{1}) \equiv \frac{1}{e^{\beta\hbar\omega^{\rm sq}_{1}}-1}$, can acquire negative occupations, meaning unphysical contributions to the path integral.
Indeed, the definition of the squeezed-field path integral (\ref{eq: definition_partition}) does not allow the fields $b_{\bm k}$ to occupy redundant states with $1 < |v_{\bm k}|^2$.
Thus, in order to describe the NG-mode branch, we need to take higher-order vertices of the HP expansion into account beyond the linear approximation.

However, we note that the NG modes are gapped out to high-energy plasmon modes in real solids because of the coupling to the electromagnetic gauge field, i.e., the Anderson--Higgs mechanism~\cite{altland2010condensed,higgs1964broken}.
As we will see below, the linearized Lagrangian leads to the dispersion relation of the Higgs mode in the spectral functions.
The branch of the Higgs modes exists as low-energy states with or without the coupling to the gauge field~\cite{altland2010condensed,higgs1964broken}.
Therefore, we expect that our approach can be applied to analyzing dynamical properties of the Higgs mode in real BCS superconductors.

\section{Observables} \label{sec: applications}

In this section, we apply the squeezed-field path-integral formalism to experimentally relevant quantities.
Specifically, we discuss the properties of the single-particle spectral function~\cite{tsuchiya2009single}.

\begin{figure*}
\begin{center}
\includegraphics[width=180mm]{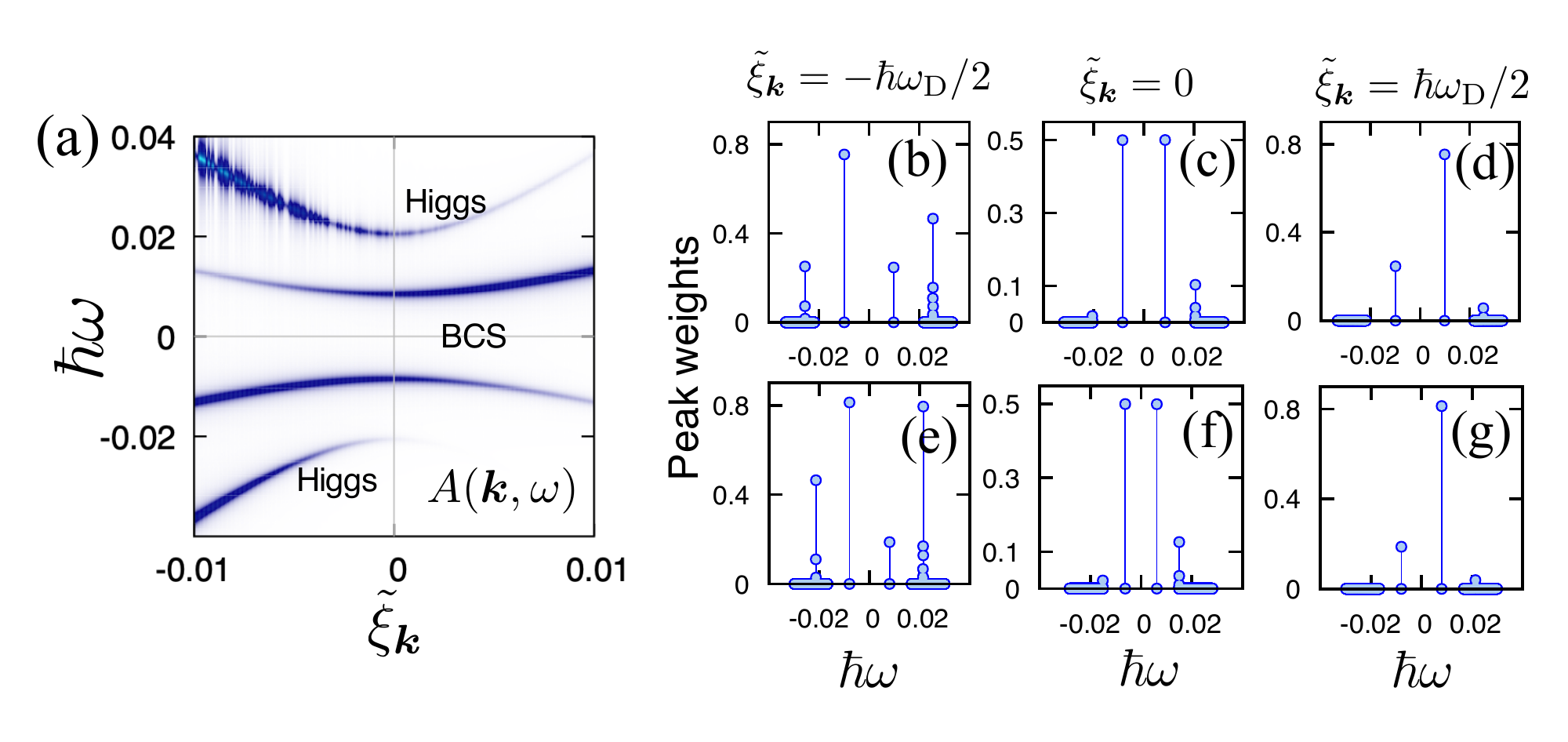}
\vspace{-8mm}
\caption{
(a): The single-particle spectral function $A({\bm k},\omega)$ calculated via the squeezed-field path integral for the BCS Hamiltonian.
We assume $gN_{\rm F}=1.0$, $\hbar\omega_{\rm D}=0.01E_{\rm F}$, $T=0$, and $N=400$.
The vertical and horizontal axes are measured in units of $E_{\rm F}$.
To plot the spectral function, we utilize the Lorentz function $P_{\rm L}(\omega)=\frac{1}{\pi}\frac{\delta}{\omega^2+\delta^2}$ and $\delta = 6.0\times 10^{-4}$ in units of the Fermi energy $E_{\rm F}$.
The peaks are rescaled by multiplying $P_{L}(\omega)$ by a small regularization factor $\pi \delta$.
We emphasize that the broadened contrast of the Higgs branch for the hole regime ${\tilde \xi}_{\bm k}<0$ is due to the artificial choice of the finite width of the Lorentz function, and that the energy levels are continuously distributed around the dispersion.
(b-d):
The spectral peak weights of $A({\bm k},\omega)$ for $gN_{\rm F}=1.0$ and $N=1000$.
The panels (b), (c), and (d) correspond to ${\tilde \xi}_{\bm k}=-\hbar\omega_{\rm D}/2$, ${\tilde \xi}_{\bm k}=0$, and ${\tilde \xi}_{\bm k}=\hbar\omega_{\rm D}/2$, respectively.
(e-g):
The peak weights for $gN_{\rm F}=0.8$ and $N=1000$, and at (e) ${\tilde \xi}_{\bm k}=-\hbar\omega_{\rm D}/2$, (f) ${\tilde \xi}_{\bm k}=0$, and (g) ${\tilde \xi}_{\bm k}=\hbar\omega_{\rm D}/2$.
}
\label{fig: spectral_function}
\end{center}
\end{figure*}

To obtain the single-particle spectral function, we consider the single-particle Green's function in the imaginary time axis~\cite{fetter2012quantum}
\begin{align}
G({\bm k},\tau-\tau') = -\left\langle T_{\tau}\left[{\hat c}_{{\bm k},\uparrow}(\tau){\hat c}^{+}_{{\bm k},\uparrow}(\tau')\right]\right\rangle.
\end{align}
where $T_{\tau}\{\cdots\}$ is the time-ordering operation for the Heisenberg field operators ${\hat c}_{{\bm k},\uparrow}(\tau) \equiv e^{\tau {\hat H}/\hbar}{\hat c}_{{\bm k},\uparrow}e^{-\tau {\hat H}/\hbar}$ and ${\hat c}^{+}_{{\bm k},\uparrow}(\tau) \equiv e^{\tau {\hat H}/\hbar}{\hat c}^{\dagger}_{{\bm k},\uparrow}e^{-\tau {\hat H}/\hbar}$.
The Matsubara frequency expansion of finite-temperature quantum-field theory \cite{fetter2012quantum} decomposes $G({\bm k},\tau-\tau')$ into the Fourier components ${\tilde G}({\bm k},i\omega_{n})$ with $\omega_{n}=\pi(2n+1)k_{\rm B}T$ for $n \in \mathbb{Z}$.
The analytical continuation of ${\tilde G}({\bm k},i\omega_{n})$ from the imaginary axis to the real axis of the complex plane leads to the spectral function of single-particle excitations, see, e.g., Ref.~\cite{tsuchiya2009single}:
\begin{align}
A({\bm k},\omega) = -\frac{1}{\pi}{\rm Im}\left[{\tilde G}({\bm k},i\omega_{n}\rightarrow \omega + i \delta)\right],
\end{align}
where $\delta>0$ is infinitesimally small.
In the Bogoliubov approximation, the single-particle spectral function $A({\bm k},\omega)$ reduces to a sum of two $\delta$-functions at $\omega = \omega_{\bm k}$ and at $\omega=-\omega_{\bm k}$, corresponding to the quasi-particle and quasi-hole excitations of the superconductor, respectively~\cite{fetter2012quantum}:
\begin{align}
A({\bm k},\omega) 
&\approx {\overline u}^2_{\bm k}\delta(\omega-\omega_{\bm k}) + {\overline v}^2_{\bm k}\delta(\omega+\omega_{\bm k}) \\
&\equiv A_{\rm MF}({\bm k},\omega). \nonumber 
\end{align}
The spectral weight functions ${\overline u}^2_{\bm k}$ and ${\overline v}^2_{\bm k}$ are given by the values of the mean-field ground state from the minimization of the BCS energy functional.
We note that perturbative evaluations of this spectral function for attractive BCS-type models beyond the BCS approximation have been reported in the literature~\cite{tsuchiya2009single,watanabe2013low}.

We discuss a higher-order correction to the mean-field result of the single-particle spectral function, which arises due to the order-parameter fluctuations of the squeezing fields $v_{\bm k}$.
We assume once again that the fluctuations of the fields $v_{\bm k}$ are small, to approximate the squeezed-field Lagrangian at quadratic order, via ${\cal L}^{(2)}_{\rm B}$, see Eq.~(\ref{eq: lagrangian_quad_boson}).
We assume that the system is at zero temperature.
Within the linear approximation, the quasiparticles and the squeezing modes are not explicitly correlated to each other.
Therefore, we can utilize the Wick theorem of Gaussian functional integrations~\cite{altland2010condensed}.
The Wick theorem states that $G({\bm k},\tau) \approx - \langle u_{\bm k}(\tau) u_{\bm k}(0) \rangle_{\rm sq} \langle \psi_{{\bm k},\uparrow}(\tau) {\overline \psi}_{{\bm k},\uparrow}(0) \rangle_{\rm sq} - \langle v_{\bm k}(\tau) v^*_{\bm k}(0) \rangle_{\rm sq} \langle {\overline \psi}_{-{\bm k},\downarrow}(\tau) \psi_{{\bm k},\downarrow}(0) \rangle_{\rm sq}$.
Note that $\langle \cdots \rangle_{\rm sq}$ implies the functional average with the squeezed-field path integral of the quadratic-order Lagrangian.
Utilizing the results of Appendix~\ref{app: green_function}, we obtain the spectral function with the fluctuation corrections 
\begin{align}
A({\bm k},\omega) 
&\approx {\overline u}^2_{\bm k}\delta(\omega-\omega_{\bm k}) + {\overline v}^2_{\bm k}\delta(\omega+\omega_{\bm k}) \nonumber \\
&\;\;\;\; + \frac{{\overline v}^2_{\bm k}}{4{\overline u}^2_{\bm k}}\sum_{s=2}^{N}|{\cal U}_{{\bm k},s}+{\cal V}_{{\bm k},s}|^2\delta(\omega-\omega_{\bm k}-\omega_{s}^{\rm sq}) \nonumber \\
&\;\;\;\; + \sum_{s=2}^{N}|{\cal V}_{{\bm k},s}|^2\delta(\omega+\omega_{\bm k}+\omega^{\rm sq}_{s}). \label{eq: spectral_squeezed}
\end{align}
The first two $\delta$-functions are the mean-field contributions of $A_{\rm MF}({\bm k},\omega)$.
We find that the correction terms to $A_{\rm MF}({\bm k},\omega)$ give sideband peaks to the total spectrum at $\omega = \omega_{\bm k}+\omega^{\rm sq}_{s}$ and at $\omega = -\omega_{\bm k}-\omega^{\rm sq}_{s}$, respectively, reflecting the fluctuations of the order parameter.
The spectral weights of the sideband peaks are determined by the transformation matrices of the bosonic Bogoliubov transformation ${\cal U}$ and ${\cal V}$, which have been introduced in Sec.~\ref{sec: spectrum}.

We note that in the expression (\ref{eq: spectral_squeezed}) we have not included the zero-mode contributions of $H_{\rm B}^{(2)}$.
If we evaluate these within the linear approximation, we find that the zero modes change the spectral weight of the quasiparticle at $\omega = \omega_{\bm k} - \omega^{\rm sq}_{1} = \omega_{\bm k} + 0^{+}$ and that of the quasihole at $\omega = -\omega_{\bm k}+\omega^{\rm sq}_{1}=-\omega_{\bm k}-0^+$, see also Appendix~\ref{app: green_function}.
However, as discussed in Ref.~\cite{lewenstein1996quantum}, the zero-momentum and zero-energy Bogoliubov mode describes the momentum of the condensed state, rather than an intrinsic excitation. 
Furthermore, as mentioned above, in a real solid this NG mode is not gapless, but has a non-zero energy which is the plasmon energy. 
For this reason we have not included this contribution in Eq.~(\ref{eq: spectral_squeezed}).
In the following, we focus on the finite-energy contributions that give the properties of the Higgs mode.
Ignoring the zero-energy contributions does not affect the following consequences on the Higgs mode.

In Fig.~\ref{fig: spectral_function}(a), we plot Eq.~(\ref{eq: spectral_squeezed}) as a function of $\omega$ and ${\tilde \xi}_{\bm k}=\epsilon_{\bm k}-\mu' = \epsilon_{\bm k} - E_{\rm F}$.
The peak contrast of $A({\bm k},\omega)$ indicates that there are two peaks above and below the peaks of the quasiparticle and quasihole excitations of the BCS approximation. 
These high-energy peaks are the Higgs amplitude mode, which are intuitively visualized as the fluctuation modes of the amplitude of the order-parameter field $\Delta = \sum_{\bm k}u_{\bm k}v_{\bm k}$~\cite{pekker2015amplitude}.
In Figs.~\ref{fig: spectral_function}(c) and (f), we display the spectral peak weights of $A({\bm k},\omega)$ near the Fermi surface, i.e., around ${\tilde \xi}_{\bm k}=0$.
There, the weights of the Higgs mode are sufficiently small compared to the weight of the quasiparticle excitation, which is given by ${\overline u}_{\bm k}^2$.
This supports that the HP expansion of $u_{\bm k}$ is justified near the Fermi surface in describing the corrections to the mean-field result.
In addition, the particle regime for $\xi_{\bm k}' \gtrsim \hbar\omega_{\rm D}/2$ is also a valid regime of the expansion, see Figs.~\ref{fig: spectral_function}(d) and (g).

\begin{figure}
\begin{center}
\includegraphics[width=90mm]{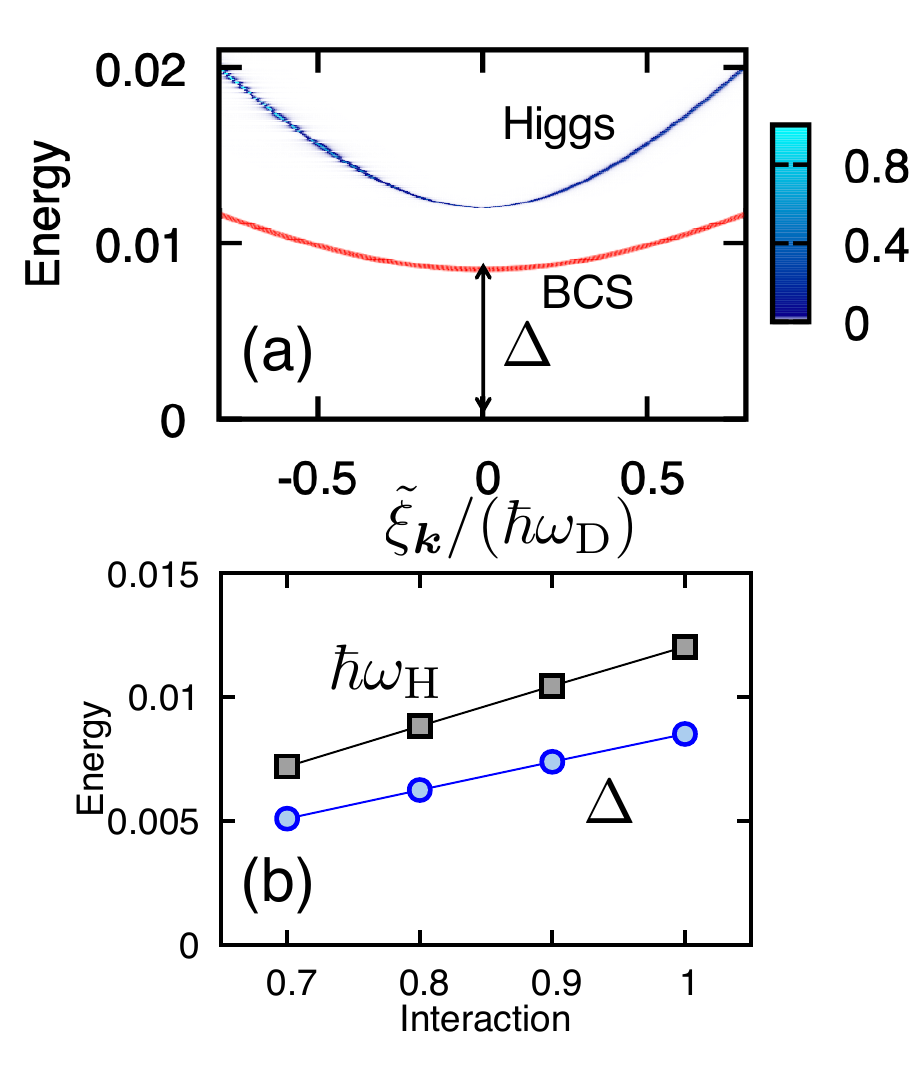}
\vspace{-8mm}
\caption{
(a): Comparison between the Higgs dispersion (blue) and the BCS dispersion (red). 
The matrix size for this plot is $N=400$ and the interaction strength is $gN_{\rm F}=1.0$.
The color bar indicates the magnitude of $|{\cal U}_{{\bm k},s}+{\cal V}_{{\bm k},s}|^2$.
The horizontal axis is ${\tilde \xi}_{\bm k}/(\hbar\omega_{\rm D})$.
The vertical axis is measured in units of $E_{\rm F}$.
(b): Comparison between the Higgs gap $\hbar\omega_{\rm H}$ (black square) and the gap function $\Delta$ (blue circle).
The horizontal axis is $gN_{\rm F}$.
The energies are measured in units of $E_{\rm F}$.
The matrix size is $N=1000$.
}
\label{fig: higgs_mass}
\end{center}
\end{figure}

However, in the hole regime for $\xi_{\bm k}' \lesssim -\hbar\omega_{\rm D}/2$ [Figs.~\ref{fig: spectral_function}(b) and (d)], the value of ${\overline v}_{\bm k}^2$ is close to unity, and the sideband peaks have larger values than ${\overline u}^2_{\bm k}$.
Therefore, the dispersion relation of the Higgs mode of this regime is only qualitatively correct within the linear approximation of the squeezed-field Lagrangian.
Furthermore, the large values of the squeezing-field corrections also imply that the Higgs dispersion in this regime may be significantly modified by higher-order nonlinear corrections of the vertices.
We note that modifications of the sideband curvature of the squeezing mode due to higher-order nonlinear couplings have been discussed in previous works~\cite{seifie2019squeezed,nagao2020fluctuations}.

The information of the dispersion relation of the Higgs mode is embedded in the Bogoliubov matrices ${\cal U}=({\cal U}_{{\bm k},s})$ and ${\cal V}=({\cal V}_{{\bm k},s})$.
In Fig.~\ref{fig: higgs_mass}(a), we display $|{\cal U}_{{\bm k},s}+{\cal V}_{{\bm k},s}|^2$ as a function of ${\tilde \xi}_{\bm k}$ and $\hbar\omega_{\rm s}$ for $gN_{\rm F}=1.0$, $\hbar\omega_{\rm D}=0.01E_{\rm F}$, and $N=400$.
The maximum value of $|{\cal U}_{{\bm k},s}+{\cal V}_{{\bm k},s}|^2$ for each ${\bm k}$ forms a clear band dispersion with an energy gap $\hbar\omega_{\rm H}$ at the Fermi surface.
We find that the Higgs dispersion has a larger energy gap than that of the quasiparticle dispersion predicted by the BCS approximation, for these parameters.
Moreover, the dispersion relation of the Higgs mode has a larger curvature than that of the quasiparticle dispersion.
We note that the band width of the Higgs-mode branch stems from the width of the continuous distribution of $\lambda_{s}$, as shown in Fig.~\ref{fig: eigenvalue}.
We also note that, if we plot $|{\cal U}_{{\bm k},s}|^2$ or $|{\cal V}_{{\bm k},s}|^2$ instead of $|{\cal U}_{{\bm k},s}+{\cal V}_{{\bm k},s}|^2$, the resulting band dispersion is the same as that of $|{\cal U}_{{\bm k},s}+{\cal V}_{{\bm k},s}|^2$.

In Fig.~\ref{fig: higgs_mass}(b), we display the extracted Higgs gap $\hbar\omega_{\rm H}$ as a function of $gN_{\rm F}$.
For all $gN_{\rm F}$, we find that the Higgs gap $\hbar\omega_{\rm H}$ is larger than the mean-field gap function $\Delta$, but smaller than $2\Delta$, implying an inequality $\Delta < \hbar\omega_{\rm H} < 2\Delta$, for this regime and for this level of approximation.
In the mean-field analysis of the BCS Hamiltonian, the Higgs gap is derived to be twice the superconducting gap: $\hbar\omega^{\rm MFA}_{\rm H} = 2\Delta$~\cite{tsuji2015theory,pekker2015amplitude}.
Hence, our numerical result displays a reduction of the Higgs gap from the mean-field Higgs gap.
This reduction is due to a combined effect of the quantum fluctuations of the squeezing fields, and the nonzero correlations between the different momentum modes within the linear approximation.

\section{Conclusions} \label{sec: conclusions}

We have developed a squeezed-field path integral description of BCS superconductors. Utilizing a completeness relation of squeezed-fermionic coherent states, we have constructed a generalized Grassmann path integral, that describes collective excitations of the order parameter and quasi-particle excitations in a single framework. The collective excitations of the order parameter are described by the dynamics of the coupled squeezing parameters of the system, which are naturally expressed as a classical Anderson pseudospin model.

In particular, we have demonstrated that the eigenmodes of the squeezing sector of the path integral  describe the Higgs mode of the superconductor. We determine the Higgs energy gap and the Higgs dispersion via numerical evaluation of the eigenmodes of the squeezing sector. The resulting Higgs spectrum is reflected in the single-particle spectral function, which displays sidepeaks of the quasi-particle peaks.

In this discussion we have ignored the Nambu--Goldstone mode that is predicted by the BCS model. 
The BCS model predicts this gapless excitation mode, in contrast to the gapped plasmon mode observed in solids, and predicted by the Anderson-Higgs mechanism. 
We will include our description to capture this feature of superconductors elsewhere.

In conclusion, we emphasize that the formalism that we have  put forth in this paper advances the theoretical description of superconductors in a fundamental way, and gives a new perspective on collective excitations, such as the Higgs mode, and their coupling to quasi-particle excitations of the superconductor. 
Furthermore we note that the formalism that we have developed here, lays out a framework that can be applied not only to a wide range of superconducting states, such as of higher orbital symmetry or of topological nature, but also to any ordered state that emerges in a fermionic system and can be described by an order parameter composed of two fermions. 
Therefore our formalism enables new insight into a broad range of physical systems that are central to many-body theory.

\begin{acknowledgments}

This work was supported by the Deutsche Forschungsgemeinschaft (DFG) through the SFB 925 and the Cluster of Excellence `Advanced Imaging of Matter' of the DFG EXC 2056 - project ID 390715994.
D.L. acknowledges financial support under a Katholischer Akademischer Auslandsdienst (KAAD) stipend.
We thank Ilias M. H. Seifie and 	Caroline Nowoczyn for interesting and fruitful discussions.

\end{acknowledgments}

\appendix 

\clearpage

\vspace{5cm}

\section{Evaluation of the path-integral overlaps} \label{app: evaluation_overlap}

We write $|\Psi_{j}\rangle \equiv {\hat S}({\bm \Omega}_{j})|{\bm \psi}_{j}\rangle$ as the squeezed-fermionic coherent state at time $\tau = j \Delta \tau$.
The overlaps of the squeezed-field Grassmann path integrals are given by
\begin{align}
 \langle \Psi_{j} | \Psi_{j-1} \rangle = \langle {\bm \psi}_{j} | {\hat S}^{\dagger}({\bm \Omega}_{j}) {\hat S}({\bm \Omega}_{j-1}) | {\bm \psi}_{j-1} \rangle.
\end{align}
We perform an expansion ${\hat S}^{\dagger}({\bm \Omega}_{j}) {\hat S}({\bm \Omega}_{j-1}) = {\hat 1} + \Delta \tau \delta {\hat S}({\bm \Omega}_{j}) + \cdots$ to obtain 
\begin{align}
\delta {\hat S}({\bm \Omega}_{j}) = &-\sum_{\bm k}{\dot \theta}_{{\bm k},j}{\hat S}^{\dagger}({\bm \Omega}_{j})\partial_{\theta_{{\bm k},j}}{\hat S}({\bm \Omega}_{j}) \nonumber \\
&-\sum_{\bm k}{\dot \varphi}_{{\bm k},j}{\hat S}^{\dagger}({\bm \Omega}_{j})\partial_{\varphi_{{\bm k},j}}{\hat S}({\bm \Omega}_{j}).
\end{align}
One can prove that 
\begin{align}
{\hat S}^{\dagger}({\bm \Omega}_{j})\partial_{\theta_{{\bm k},j}}{\hat S}({\bm \Omega}_{j}) &= \frac{e^{i\varphi_{{\bm k},j}}}{2}{\hat c}^{\dagger}_{{\bm k},\uparrow}{\hat c}^{\dagger}_{-{\bm k},\downarrow} - \frac{e^{-i\varphi_{{\bm k},j}}}{2}{\hat c}_{-{\bm k},\downarrow}{\hat c}_{{\bm k},\uparrow}, \\
{\hat S}^{\dagger}({\bm \Omega}_{j})\partial_{\varphi_{{\bm k},j}}{\hat S}({\bm \Omega}_{j}) &= \frac{i}{2}{\hat S}^{\dagger}({\bm \Omega}_{j})({\hat n}_{{\bm k},\uparrow}+{\hat n}_{-{\bm k},\downarrow}) {\hat S}({\bm \Omega}_{j}) \nonumber \\
&\;\;\;\;\; - \frac{i}{2}({\hat n}_{{\bm k},\uparrow}+{\hat n}_{-{\bm k},\downarrow}).
\end{align}
See also Ref.~\cite{nagao2020fluctuations}.
These relations lead to 
\begin{widetext}
\begin{align}
 \langle \Psi_{j} | \Psi_{j-1} \rangle 
 &= \langle {\bm \psi}_{j} | {\bm \psi}_{j-1} \rangle e^{-\Delta \tau  \sum_{\bm k}\frac{1}{2}{\dot \theta}_{{\bm k},j}\left( e^{i\varphi_{{\bm k},j}}{\overline \psi}_{{\bm k},\uparrow}{\overline \psi}_{-{\bm k},\downarrow} - e^{-i\varphi_{{\bm k},j}}\psi_{-{\bm k},\downarrow}\psi_{{\bm k},\uparrow} \right) } \nonumber \\
 &\times e^{- \Delta \tau \sum_{\bm k} i {\dot \varphi}_{{\bm k},j}\left( |v_{{\bm k},j}|^2 - |v_{{\bm k},j}|^2{\overline \psi}_{{\bm k},\uparrow}\psi_{{\bm k},\uparrow} - |v_{{\bm k},j}|^2{\overline \psi}_{-{\bm k},\downarrow}\psi_{-{\bm k},\downarrow} + u_{{\bm k},j}v_{{\bm k},j}{\overline \psi}_{{\bm k},\uparrow}{\overline \psi}_{-{\bm k},\downarrow} + u_{{\bm k},j}v^*_{{\bm k},j}\psi_{-{\bm k},\downarrow}\psi_{{\bm k},\uparrow} \right)} + {\cal O}(\Delta \tau^2), \\
 \langle {\bm \psi}_{j} | {\bm \psi}_{j-1} \rangle 
 &= \exp\left( \frac{\Delta \tau}{2}\sum_{{\bm k},\sigma}\frac{{\overline \psi_{{\bm k},\sigma}}(\tau_{j})-{\overline \psi_{{\bm k},\sigma}}(\tau_{j-1})}{\Delta \tau}\psi_{{\bm k},\sigma}(\tau_{j-1}) - \frac{\Delta \tau}{2}\sum_{{\bm k},\sigma}{\overline \psi_{{\bm k},\sigma}}(\tau_{j})\frac{\psi_{{\bm k},\sigma}(\tau_{j})-\psi_{{\bm k},\sigma}(\tau_{j-1})}{\Delta \tau} \right).
\end{align}
\end{widetext}

\section{$2N$-dimensional Bogoliubov matrix} \label{app: bogoliubov}

The bosonic Hamiltonian $H^{(2)}_{\rm B}$ can be expressed as
\begin{align}
H^{(2)}_{\rm B} &=
\sum_{i=1}^{N}
\begin{pmatrix}
b^*_{i} & b_{i}
\end{pmatrix}
\begin{pmatrix}
A_{ii} & B_{ii} \\
B_{ii} & A_{ii} 
\end{pmatrix}
\begin{pmatrix}
b_{i} \\
b^*_{i}
\end{pmatrix} \nonumber \\
&+
\sum_{i_1 \neq i_2}
\begin{pmatrix}
b^*_{i_1} & b_{i_1}
\end{pmatrix}
\begin{pmatrix}
A_{i_1i_2} & B_{i_1i_2} \\
B_{i_1i_2} & A_{i_1i_2} 
\end{pmatrix}
\begin{pmatrix}
b_{i_2} \\
b^*_{i_2}
\end{pmatrix}
.
\end{align}
Using these matrices $A$ and $B$, the Bogoliubov matrix $M$ is written as 
\begin{align}
M = 
\begin{pmatrix}
A_{11} & A_{12} & \cdots & B_{11} & B_{12} & \cdots \\ 
A_{21} & A_{22} & \cdots & B_{21} & B_{22} & \cdots \\ 
\vdots & \vdots & & \vdots & \vdots \\
B_{11} & B_{12} & \cdots & A_{11} & A_{12} & \cdots  \\ 
B_{21} & B_{22} & \cdots & A_{21} & A_{22} & \cdots  \\
\vdots & \vdots & & \vdots & \vdots 
\end{pmatrix}
.
\end{align}
Each matrix element is defined as follows:
\begin{align}
A_{ii} 
&= \left[ \xi_{i} + E_{\rm F}-\mu' + \frac{{\overline v}_{i}}{{\overline u}_{i}}\frac{\Delta}{2} + \frac{\Delta}{4}\frac{{\overline v}^3_{i}}{{\overline u}^3_{i}} \right] - \frac{gN_{\rm F}\Delta \xi}{4}\frac{{\overline v}^4_{i}}{{\overline u}^2_{i}} \nonumber \\
&\;\;\;\;\;\;\;\; - \frac{gN_{\rm F}\Delta \xi}{2}, \nonumber \\
B_{ii} 
&= \frac{\Delta}{4}\frac{{\overline v}_{i}^3}{{\overline u}_{i}^3} - \frac{gN_{\rm F}\Delta \xi}{2} {\overline v}^2_{i} - \frac{gN_{\rm F}\Delta \xi}{4}\frac{{\overline v}^4_{i}}{{\overline u}^2_{i}}, \nonumber \\
A_{ij} &= -gN_{\rm F}\Delta \xi {\overline v}_{i} {\overline v}_{j} - \frac{gN_{\rm F}\Delta \xi}{4}\frac{ {\overline v}_{i}^2}{ {\overline u}_{i}} \frac{ {\overline v}_{j}^2}{ {\overline u}_{j}}  - \frac{gN_{\rm F}\Delta \xi}{2}{\overline u}_{i}{\overline u}_{j} \nonumber \\
&\;\;\;\;\;\;\;\;\;\;\;\;\;\;\;\;\;\; + \frac{gN_{\rm F}\Delta \xi}{4}\left[\frac{{\overline v}^2_{i}}{{\overline u}_{i}} {\overline u}_{j}  + \frac{{\overline v}^2_{j}}{{\overline u}_{j}} {\overline u}_{i} \right],\;\; \text{for}\; i\neq j, \nonumber \\
B_{ij} &= -gN_{\rm F}\Delta \xi {\overline v}_{i}{\overline v}_{j} - \frac{gN_{\rm F}\Delta \xi}{4} \frac{ {\overline v}_{i}^2}{ {\overline u}_{i}} \frac{ {\overline v}_{j}^2}{ {\overline u}_{j}} \nonumber \\
&\;\;\;\;\;\;\;\;\;\;\;\;\;\;\;\;\;\; + \frac{gN_{\rm F}\Delta \xi}{4}\left[\frac{{\overline v}^2_{i}}{{\overline u}_{i}} {\overline u}_{j}  + \frac{{\overline v}^2_{j}}{{\overline u}_{j}} {\overline u}_{i} \right],\;\; \text{for}\; i\neq j, \nonumber
\end{align}
where $\xi_{i} \in [-\hbar\omega_{\rm D},\hbar\omega_{\rm D}]$.

\section{The imaginary-time Green's function within the linear approximation} \label{app: green_function}

For the linearized path-integral action, the imaginary-time Green's function reduces to 
\begin{align}
G({\bm k},\tau) \approx 
&-\langle u_{{\bm k}}(\tau) u_{{\bm k}}(0) \rangle_{\rm sq} \langle \psi_{{\bm k},\uparrow}(\tau){\overline \psi}_{{\bm k},\uparrow}(0) \rangle_{\rm sq} \nonumber \\
&- \langle v_{{\bm k}}(\tau) v^*_{{\bm k}}(0) \rangle_{\rm sq} \langle {\overline \psi}_{-{\bm k},\downarrow}(\tau) \psi_{-{\bm k},\downarrow}(0) \rangle_{\rm sq}.
\end{align}
We expand $u_{\bm k}(\tau)$ in $b_{\bm k}(\tau) = v_{\bm k}(\tau) - {\overline v}_{\bm k}$ to obtain the leading order correction to the BCS mean-field result, i.e. 
\begin{align}
G({\bm k},\tau) 
&\approx -{\overline u}^2_{\bm k}[1-f_{\rm F}(\omega_{\bm k})]e^{-\omega_{\bm k}\tau} - {\overline v}^2_{\bm k}f_{\rm F}(\omega_{\bm k})e^{\omega_{\bm k}\tau}   \nonumber \\
&\;\;\;\;+ G_{1}({\bm k},\tau) + G_{2}({\bm k},\tau) + G_{3}({\bm k},\tau) + G_{4}({\bm k},\tau).
\end{align}
For this equation, we have defined 
\begin{align}
G_{1}({\bm k},\tau) 
&= - \frac{{\overline v}_{\bm k}^2}{4{\overline u}_{\bm k}^2}[1-f_{\rm F}(\omega_{\bm k})]\sum_{s=1}^{N}|{\cal U}_{{\bm k},s}+{\cal V}_{{\bm k},s}|^2 \nonumber \\
&\;\;\;\;\;\;\;\;\;\;\;\;\;\; \times [1+f_{\rm B}(\omega^{\rm sq}_{s})]e^{-\tau(\omega_{s}^{\rm sq}+\omega_{\bm k})}, \\
G_{2}({\bm k},\tau) &= - \frac{{\overline v}_{\bm k}^2}{4{\overline u}_{\bm k}^2}[1-f_{\rm F}(\omega_{\bm k})]\sum_{s=1}^{N}|{\cal U}_{{\bm k},s}+{\cal V}_{{\bm k},s}|^2 \nonumber \\
&\;\;\;\;\;\;\;\;\;\;\;\;\;\; \times f_{\rm B}(\omega^{\rm sq}_{s}) e^{-\tau(\omega_{\bm k}-\omega_{s}^{\rm sq})}, \\
G_{3}({\bm k},\tau) &=
- f_{\rm F}(\omega_{\bm k})\sum_{s=1}^{N}|{\cal U}_{{\bm k},s}|^2 [1+f_{\rm B}(\omega^{\rm sq}_{s})] e^{\tau(\omega_{\bm k}-\omega_{s}^{\rm sq})}, \\
G_{4}({\bm k},\tau) &=
- f_{\rm F}(\omega_{\bm k})\sum_{s=1}^{N}|{\cal V}_{{\bm k},s}|^2 f_{\rm B}(\omega^{\rm sq}_{s}) e^{\tau(\omega_{\bm k}+\omega_{s}^{\rm sq})}.
\end{align}
Notice that ${\cal U}$ and ${\cal V}$ can be assumed to be real for the matrix $\Sigma_{2N}M$, because it has no imaginary term.
The free propagators of each field at finite temperatures can be calculated~\cite{altland2010condensed} as 
\begin{align}
\langle \psi_{{\bm k},\uparrow}(\tau){\overline \psi}_{{\bm k},\uparrow}(0) \rangle_{\rm sq} &= [1-f_{\rm F}(\omega_{\bm k})]e^{-\tau\omega_{\bm k}}, \\
\langle {\overline \psi}_{-{\bm k},\downarrow}(\tau) \psi_{-{\bm k},\downarrow}(0) \rangle_{\rm sq} &= f_{\rm F}(\omega_{\bm k})e^{\tau\omega_{\bm k}}, \\
\langle \beta_{s}(\tau) \beta^{*}_{s}(0) \rangle_{\rm sq} &= [1+f_{\rm B}(\omega^{\rm sq}_{s})]e^{-\tau\omega^{\rm sq}_{s}},\\
\langle \beta^{*}_{s}(\tau) \beta_{s}(0) \rangle_{\rm sq} &= f_{\rm B}(\omega^{\rm sq}_{s})e^{\tau\omega^{\rm sq}_{s}}.
\end{align}
$f_{\rm B/F}(\omega)=\frac{1}{e^{\beta \hbar \omega} \mp 1}$ is the Bose or Fermi distribution function.

The Matsubara-Fourier transforms of $G_{1}$, $G_{2}$, $G_{3}$, and $G_{4}$ with $\omega_{n}=\pi(2n+1)/(\hbar\beta)$ read
\begin{align}
{\tilde G}_{1}({\bm k},i\omega_{n}) &= \frac{{\overline v}_{\bm k}^2}{4{\overline u}_{\bm k}^2}\sum_{s=1}^{N}\frac{|{\cal U}_{{\bm k},s}+{\cal V}_{{\bm k},s}|^2 \Theta_{1}(\omega_{\bm k},\omega^{\rm sq}_{s})}{i\omega_{n}-(\omega_{\bm k}+\omega_{s}^{\rm sq})}, \\
{\tilde G}_{2}({\bm k},i\omega_{n}) &= \frac{{\overline v}_{\bm k}^2}{4{\overline u}_{\bm k}^2}\sum_{s=1}^{N}\frac{|{\cal U}_{{\bm k},s}+{\cal V}_{{\bm k},s}|^2 \Theta_{2}(\omega_{\bm k},\omega^{\rm sq}_{s})}{i\omega_{n}-(\omega_{\bm k}-\omega_{s}^{\rm sq})}, \\
{\tilde G}_{3}({\bm k},i\omega_{n}) &= \sum_{s=1}^{N}\frac{|{\cal U}_{{\bm k},s}|^2\Theta_{3}(\omega_{\bm k},\omega^{\rm sq}_{s})}{i\omega_{n}+(\omega_{\bm k}-\omega^{\rm sq}_{s})}, \\
{\tilde G}_{4}({\bm k},i\omega_{n}) &= \sum_{s=1}^{N}\frac{|{\cal V}_{{\bm k},s}|^2\Theta_{4}(\omega_{\bm k},\omega^{\rm sq}_{s})}{i\omega_{n}+(\omega_{\bm k}+\omega^{\rm sq}_{s})}.
\end{align}
Notice that $e^{i\beta\hbar\omega_{n}}=e^{\pi(2n+1)}=-1$.
The functions $\Theta_{i}(\omega_{\bm k},\omega^{\rm sq}_{s})$ for each Green's function are defined by
\begin{align}
\Theta_{1}(\omega_{\bm k},\omega^{\rm sq}_{s}) &= [1-f_{\rm F}(\omega_{\bm k})][1+f_{\rm B}(\omega^{\rm sq}_{s})][1+e^{-\beta\hbar(\omega_{\bm k}+\omega^{\rm sq}_{s})}], \nonumber \\
\Theta_{2}(\omega_{\bm k},\omega^{\rm sq}_{s}) &= [1-f_{\rm F}(\omega_{\bm k})]f_{\rm B}(\omega^{\rm sq}_{s})[1+e^{-\hbar\beta(\omega_{\bm k}-\omega^{\rm sq}_{s})}], \nonumber \\
\Theta_{3}(\omega_{\bm k},\omega^{\rm sq}_{s}) &= f_{\rm F}(\omega_{\bm k})[1+f_{\rm B}(\omega^{\rm sq}_{s})][1+e^{\hbar\beta(\omega_{\bm k}-\omega^{\rm sq}_{s})}], \nonumber \\
\Theta_{4}(\omega_{\bm k},\omega^{\rm sq}_{s}) &= f_{\rm F}(\omega_{\bm k})f_{\rm B}(\omega^{\rm sq}_{s})[1+e^{\beta\hbar(\omega_{\bm k}+\omega^{\rm sq}_{s})}]. \nonumber
\end{align}

The zero-temperature limits of $\Theta_{i}(\omega_{\bm k},\omega^{\rm sq}_{s})$ depend on the sign of $\hbar\omega^{\rm sq}_{s}$.
For example, $\Theta_{4}(\omega_{\bm k},\omega^{\rm sq}_{s})$ becomes unity for $\hbar\omega^{\rm sq}_{s} > 0 $ in the zero-temperature limit, i.e., 
\begin{align}
\Theta_{4}(\omega_{\bm k},\omega^{\rm sq}_{s})
&= \frac{1+e^{\beta \hbar (\omega_{\bm k}+\omega_{s}^{\rm sq})}}{(e^{\beta \hbar \omega_{\bm k}}+1)(e^{\beta \hbar \omega_{s}^{\rm sq}}-1)} \nonumber \\
&= \frac{e^{-\beta \hbar (\omega_{\bm k}+\omega_{s}^{\rm sq})}+1}{1+e^{-\beta \hbar \omega_{\bm k}}-e^{-\beta \hbar \omega^{\rm sq}_{s}}-e^{-\beta \hbar (\omega_{\bm k}+\omega_{s}^{\rm sq})}} \nonumber \\
&\rightarrow 1. \nonumber
\end{align}
However, for $\hbar\omega^{\rm sq}_{s} < 0 $, it becomes zero. 
Indeed, 
\begin{align}
\Theta_{4}(\omega_{\bm k},\omega^{\rm sq}_{s})
&= \frac{e^{-\beta \hbar \omega_{\bm k}}+e^{\beta \hbar \omega_{s}^{\rm sq}}}{e^{\beta \hbar \omega_{s}^{\rm sq}}+e^{\beta \hbar (\omega_{s}^{\rm sq}-\omega_{\bm k})}-1-e^{-\beta \hbar \omega_{\bm k}}} \nonumber \\
&\rightarrow 0. \nonumber
\end{align}
The dependence of the signs of $\lim_{T\rightarrow 0} \Theta_{i}(\omega_{\bm k},\omega^{\rm sq}_{s})$ is summarized as follows:
\begin{align}
\lim_{T\rightarrow 0}\Theta_{1}(\omega_{\bm k},\omega^{\rm sq}_{s}) 
&= \lim_{T\rightarrow 0}\Theta_{4}(\omega_{\bm k},\omega^{\rm sq}_{s}) = 1-\delta_{s,1},  \\
\lim_{T\rightarrow 0}\Theta_{2}(\omega_{\bm k},\omega^{\rm sq}_{s}) 
&= \lim_{T\rightarrow 0}\Theta_{3}(\omega_{\bm k},\omega^{\rm sq}_{s}) = -\delta_{s,1}.
\end{align}
Therefore, we obtain Eq.~(\ref{eq: spectral_squeezed}) in the text.

\bibliography{ref}

\begin{thebibliography}{66}
\bibitem [{not()}]{note1}%
  \BibitemOpen
  \href@noop {} {\ }\bibinfo {note} {The standard definition of the Haar
  measure for the SU(2) spin-coherent states is $d{\bm \Omega} =
  \frac{2S+1}{4\pi}d\theta \sin \theta d\varphi$, for which $S$ is the strength
  of the SU(2) spins. See
  Refs.~\cite{fradkin2013field,auerbach2012interacting}.}\BibitemShut {Stop}%
\end{thebibliography}

\end{document}